\documentstyle[12pt]{article}
\title{A global fit of $\pi\pi$ and $\pi K$ elastic  scattering in ChPT with
dispersion relations}
\author{{\large \bf A. Dobado and J. R. Pel\'aez} \\
Departamento de F\'{\i}sica Te\'orica \\
Universidad Complutense de Madrid\\
 28040 Madrid, Spain }
\begin{document}
\maketitle
\begin{abstract}
We apply the one-loop results of the $SU(3)_L\times SU(3)_R$ ChPT
suplemented with the inverse amplitude method to fit the
available experimental data on $\pi\pi$ and $\pi K$ scattering. With
esentially only three parameters we describe
accurately data corresponding to six different channels, namely $(I,J)=(0,0),
(2,0), (1,1), (1/2,0), (3/2,0)$ and $(1/2,1)$. In addition we reproduce
the first resonances of the $(1,1)$ and $(1/2,1)$ channel with the right mass
corresponding to the $\rho$ and the $K^*(892)$ particles. 
\end{abstract}
\vskip 3.0cm
FT/UCM/10/92
\newpage
\baselineskip 0.83 true cm
\textheight 20 true cm
\newpage

{\bf Introduction:}    In 1979  Weinberg [1] suggested that
it is possible to  summarize many previous current algebra results
in a phenomenological lagrangian that incorporates all the constraints
coming from the chiral symmetry of the strong interactions and QCD.
 This technique, also called Chiral Perturbation Theory, was developped some
time later to one-loop level and in great detail in a celebrated
set of papers by Gasser and Leutwyler [2]. In these works, the authors showed
how it is possible to compute many different Green functions involving
low-energy pions and kaons as functions of the lowest powers of their  momenta,
their masses  and a few phenomenological parameters. In these and in further
works it was also shown that this method provides a good parametrization of
many
low-energy experimental data.

 More recently,  the strongly interacting symmetry breaking sector of
the Standard Model [3] has also been described phenomenologically using ChPT
[4].

 In both contexts , one of the main obstacles found when one tries to apply
ChPT
to higher
 energies relies in the issue of unitarity.
ChPT, being a consistent theory, is unitary in the perturbative sense. However,
as the expansion parameters are the momenta and the Nambu-Goldstone boson
masses, perturbative unitarity breaks down, sometimes even at moderate
energies [5]. Different attempts to improve this behaviour of ChPT
 and extend the applicability to higher energies, have been proposed in the
literature. These methods include the use of unitarization procedures such
as the Pad\'e expansion [5,6], the inverse amplitude method [5], and the
explicit introduction of more  fields describing resonances [7]. All them
 improve the unitarity behavior of the ChPT expansion and provide a more
accurate description of the data although there is some controversy  about
which of them is more appropriate.

In this note we show the results of the application of the inverse amplitude
method to the ChPT one-loop computation of elastic $\pi\pi$ and $\pi K$
scattering [2,8]. The obtained amplitude converges at low energies with that
of standard ChPT but it fulfills strictly elastic unitarity, since the
contribution of the right cut in the corresponding dispersion relation is taken
into account exactly in the inverse amplitude method. Using this approach
(which incidentaly is connected with the formal [1,1] Pad\'e approximant of the
one-loop amplitudes) we will make a global fit of the data for
elastic $\pi\pi$ and $\pi K$ scattering just varying the $L_1$, $L_2$ and $L_3$
ChPT constants. The resulting values of these parameters will not be quite
different from those obtained previously, but the range of energies and the
quality of the fit will be enlarged quite amazingly.

{\bf The partial waves in ChPT:} The elastic  scattering partial
waves are defined from the corresponding isospin amplitude $T_I(s,t)$  as:
\begin{equation}
t_{IJ}(s)=\frac{1}{32 K \pi}\int_{-1}^{1}d(cos \theta) P_J(cos
\theta)T_I(s,t)
\end{equation}
where $K=2$ or $1$ depending on whether the two particles in the reaction are
identical or not.
 For elastic $\pi\pi$ scattering the possible isospin channels are
$I=0,1,2$ whilst for $\pi K$ we can have $I=1/2$ or $I=3/2$. In the first case
the isospin  amplitudes $T_I$ can be written in terms of a simple function
$A(s,t,u)$ as follows; $T_0(s,t,u)=3A(s,t,u)+A(t,s,u)+A(u,t,s),
 T_1(s,t,u)=A(s,t,u)-A(t,s,u)$ and $T_2(s,t,u)=A(s,t,u)+A(t,s,u)$. In the
second, the $I=1/2$ amplitude can be written as
$T_{1/2}(s,t,u)=(3/2)T_{3/2}(u,t,s)-(1/2)T_{3/2}(s,t,u)$. The Mandelstam
variables $s,t,u$ fulfill $s+t+u=2(M_{\alpha}^2+M_{\beta}^2)$ where we use the
notation
 $\alpha$ and $\beta=\pi$ or $K$, so that we can describe both processes with
the same general formulae.
 For $s > s_{th}=(M_{\alpha}+M_{\beta})^2$ the partial waves $t_{IJ}$ can be
parametrized as:

\begin{equation}
t_{IJ}(s)= e^{i\delta_{IJ}(s) }\sin \delta_{IJ}(s)/\sigma_{\alpha \beta}(s)
\end{equation}
where
\begin{equation}
 \sigma_{\alpha\beta}(s) = \sqrt{  (1-\frac{(M_{\alpha}+M_{\beta})^2}{s})
(1-\frac{(M_{\alpha}-M_{\beta})^2}{s})}
\end{equation}
 The
$t_{IJ}(s)$
 amplitude fulfills
the elastic unitarity condition
\begin{equation}
Imt_{IJ}=\sigma_{\alpha \beta} \mid t_{IJ} \mid ^2
\end{equation}
on the right cut.

		Using standard one-loop ChPT it is possible to compute the above
scattering amplitudes to order $p^4$ (as it is customary, any Mandelstam
variable, $M_{\pi}^2$ and $M_{K}^2$ will be considered of the order of $p^2$).
The relevant functions $A(s,t,u)$ and $T_{3/2}$ were computed in [2,8].
\begin{eqnarray}
A(s,t,u)&=&(s-M_{\pi}^2)/F_{\pi}^2+B(s,t,u)+C(s,t,u)+O(E^6)  \\  \nonumber
B(s,t,u)&=&\frac{1}{F_{\pi}^4} (
\frac {M_{\pi}^4}{18} J_{\eta\eta}^r (s) +
\frac{1}{2} ( s^2 - M^4_{\pi}) J_{\pi\pi}^r(s) + \frac {1}{8} s^2 J_{KK}^r (s)
\\
\nonumber
 &+& \frac{1}{4}( t - 2 M_{\pi}^2)^2 J_{\pi\pi}^r (t) + t(s-u) [
M_{\pi\pi}^r (t)+\frac{1}{2} M_{KK}^r (t) ]+ (t \leftrightarrow u) )\\
\nonumber
  C(s,t,u)&=&\frac{4}{F_{\pi}^4} ( (2L_1^r + L_3 )( s - 2M_{\pi}^2)^2 +
 L_2^r [ (t-2M_{\pi}^2)^2 + ( u - 2M_{\pi}^2)^2] +\\   \nonumber
 &+& (4L_4^r+2L_5^r)M_{\pi}^2(s-2M_{\pi}^2) + (8L_6^r+4L_8^r)M_{\pi}^4 ) \\
\nonumber
\end{eqnarray}
and
\begin{eqnarray}
T^{3/2}(s,t,u)&=&\frac{M_{\pi}^2+M_{K}^2-s}{2F^2_{\pi}}+T^T_4(s,t,u)+T^P_4(s,t,u)+
T^U_4(s,t,u) \\  \nonumber
 T^T_4(s,t,u)&=& \frac{1}{16F^2_{\pi}}(\mu_{\pi}[10s-7M_{\pi}^2-13M_{K}^2]  \\
\nonumber
 &+& \mu_{K}[2M_{\pi}^2+6M_{K}^2-4s]
+\mu_{\eta}[5M_{\pi}^2+7M_{K}^2-6s] ) \\   \nonumber
T^P_4(s,t,u)&=& \frac{2}{F^2_{\pi}F^2_{K}}( 4L_1^r(t-2M_{\pi}^2 )(t-2M_{K}^2 )
\\   \nonumber
&+& 2L_2^r [ (s - M_{\pi}^2-M_{K}^2)^2 + (u - M_{\pi}^2-M_{K}^2)^2] \\
\nonumber
&+& L_3^r[(u - M_{\pi}^2-M_{K}^2)^2 + (t-2M_{\pi}^2 )(t-2M_{K}^2)] \\
\nonumber
&+& 4L_4^r[t(M_{\pi}^2+M_{K}^2) - 4M_{\pi}^2M_{K}^2]\\   \nonumber
&+& 2L_5^rM_{\pi}^2(M_{\pi}^2-M_{K}^2-s)+8(2L_6^r+L_8^r)M_{\pi}^2M_{K}^2 )\\
\nonumber
T^U_4(s,t,u)&=& \frac{1}{4F^2_{\pi}F^2_{K}}(t(u-s)[2M_{\pi\pi}^r(t)+M_{KK}^r(t)
]\\   \nonumber
&+&\frac{3}{2}[(s-t)(L_{\pi K}(u)+L_{K\eta}(u)-u(M_{\pi
K}^r(u)+M_{K\eta}^r(u))) \\   \nonumber
&+& (M_{K}^2 - M_{\pi}^2)(M_{\pi K}^r(u)+M_{K \eta}^r(u))] +
J_{\pi K}^r(s)(s-M_{K}^2-M_{\pi}^2)^2 \\   \nonumber
&+& \frac{1}{2} (M_{K}^2 - M_{\pi}^2) [ K_{\pi K}(u)(5u-2M_{K}^2-2M_{\pi}^2)+
K_{K \eta}(u)(3u-2M_{K}^2-2M_{\pi}^2)] \\   \nonumber
&+&\frac{1}{8}J_{\pi K}^r(u)[ 11u^2 -
12u(M_{K}^2+M_{\pi}^2)+4(M_{K}^2+M_{\pi}^2)^2]  \\   \nonumber
&+&\frac{3}{8}J_{K \eta}^r(u) (u - \frac{3}{2}(M_{K}^2+M_{\pi}^2))^2 +
\frac{1}{2}J_{\pi\pi}^r(t)t(2t-M_{\pi}^2)  \\   \nonumber
&+&\frac{3}{4}J_{KK}^r(t)t^2+\frac{1}{2}J_{\eta\eta}^r(t)M_{\pi}^2(t-\frac{8}{9}
M_{K}^2))\\   \nonumber
\end{eqnarray}
The masses $M_{\alpha}$ and the decay constants
$F_{\alpha}$  appearing in these equations
are the physical values. The relation with the corresponding constants
appearing
in the chiral lagrangian, and the functions $\mu_{\alpha}$ can be found in [8].
The transcendental functions $M^r_{\alpha\beta}, L^r_{\alpha\beta}$ and
$J^r_{\alpha\beta}$ are defined in [2]. The first terms in the above amplitudes
reproduce the well known Weinberg low-energy theorems.  The $L^r_i$ constants
can be considered as phenomenological parameters that up to constant factors
are the renormalized coupling constants of the chiral lagrangian renormalized
conventionaly at the $m_{\eta}$ scale. Their relation with the corresponding
bare constants $L_i$ and their evolution with the renormalization scale can be
found in [2]. Using
 eq.1  and eq.5  it is possible to obtain  the corresponding partial
wave amplitudes. In the general framework of ChPT they can be obtained as a
series with increasing number of $p^2$ powers, i.e.
\begin{equation}
t_{IJ}=t_{IJ}^{(0)}+t_{IJ}^{(1)}+...
\end{equation}
 where $t_{IJ}^{(0)}$ is of
order $p^2$ and corresponds to the low-energy theorem and $t_{IJ}^{(1)}$ is
of order $p^4$.  In general, the
real part of $t_{IJ}^{(1)}$ cannot be expressed  in terms of elementary
functions but it can be computed numerically. The amplitudes in eq.5 have been
used in the literature, without further elaboration, to fit the low-energy
$\pi\pi$ and $\pi K$ scattering data [2,8,9].

{\bf Dispersion relations and the inverse amplitude method:}    One very
important point concerning the partial wave amplitudes computed from ChPT
 to one-loop is the fact that they have the
appropriate cut structure, namely the left cut and the right or unitarity cut.
However, they only fulfill the unitarity condition on the right cut in
a perturbative sense i.e.
\begin{eqnarray}
  Im t_{IJ}^{(1)} = \sigma_{\alpha \beta}\mid t_{IJ}^{(0)} \mid
^2
\end{eqnarray}

Let us show now how, with the use of dispersion
theory,  it is possible to build up a completely
unitarized amplitude for the $\pi\pi$ and $\pi K$ amplitudes
starting from the one-loop ChPT result above. Let us start writing
a three
subtracted dispersion relation for the partial wave $t_{IJ}$:
\begin{equation}
t_{IJ}(s)=C_0+C_1s+C_2s^2+
\frac{s^3}{\pi}\int_{(M_{\alpha}+M_{\beta})^2}^{\infty}\frac{Im
t_{IJ}(s')ds'}{s'^3(s'-s-i\epsilon)} + LC(t_{IJ})
\end{equation}
where
$LC(t_{IJ})$ represent the left cut contribution. In particular, this equation
is fulfilled by the standard one-loop ChPT result. Note that three subtractions
are needed to ensure the convergence of the integrals in this
approximation since the one-loop ChPT amplitudes are second order polynomials
modulo $log$ factors (higher order ChPT amplitudes require more
subtractions but we do not know how many are needed for the exact result). To
the one-loop level, the imaginary part of the amplitude can be written on the
right cut integral as $Im t_{IJ}\simeq Im t_{IJ}^{(1)}=\sigma t_{IJ}^{(0)2}$
(from the left cut there is a generic $t_{IJ}^{(1)}$ contribution).

The substraction terms appearing in the above dispersion relation can be
expanded
in powers of $M_{\alpha}^2$, and they have contributions from $t_{IJ}^{(0)}$
and
from $t_{IJ}^{(1)}$ that can be written as $a_0+a_1s$ and $b_0+b_1s+b_2s^2$ and
also from higher order terms of the ChPT series i.e $C_0=a_0+b_0+...$ and
$C_1=a_1+b_1+...$. The  $a$ and $b$ constants depend also on the
$L^r_i$ parameters. Then, in the one-loop ChPT aproximation, the exact
amplitude $t(s)$ can be written as:   \begin{equation}
t_{IJ}(s)\simeq t_{IJ}^{(0)}(s)+t_{IJ}^{(1)}(s)
\end{equation} where:
\begin{eqnarray}
t_{IJ}^{(0)} &=& a_0+a_1s    \\   \nonumber
t_{IJ}^{(1)} &=& b_0+b_1s+b_2s^2+   \\  \nonumber
&+&\frac{s^3}{\pi}\int_{(M_{\alpha}+M_{\beta})^2}^{\infty}\frac{\sigma
t_{IJ}^{(0)}(s')^2ds'}{s'^3(s'-s-i\epsilon)}+LC(t_{IJ})  \nonumber
\end{eqnarray}
where in the left cut contribution we have to integrate here
$Imt_{IJ}^{(1)}(s)$ as
an aproximation to $t_{IJ}(s)$.
 In some sense, we can understand eq.9,10 and
eq.11 in a rather different way that will be useful later: we can assume
that the dispersion relation in eq.9 is fulfilled by the exact amplitude $t(s)$
and we solve this equation approximately by introducing inside the left
and right cut integrals the one-loop prediction for $Imt(s)$ to find again
eq.10. However, we would like to stress again that this result does not fulfill
the elastic unitarity condition in eq.4 but only the perturbative version in
eq.8. In fact this happens to any order in the ChPT expansion since a
polynomial can never fulfill eq.4. However, there are other ways to use the
information contained in the ChPT series apart from the direct comparison with
the experiment of the truncated series.
				 Instead
of using the  dispersion relation for the loop ChPT amplitude
we can try a  three subtracted dispersion relation for
the inverse of $t_{IJ}$ or more exactly for the auxiliar function
$G(s)=t_{IJ}^{(0)2}/t_{IJ}$ namely:  \begin{equation}
G(s)=G_0+G_1s+G_2s^2+     \\   \nonumber
\frac{s^3}{\pi}\int_{(M_{\alpha}+M_{\beta})^2}^{\infty}\frac{ImG(s')ds'}{s'^3(s'-s-i\epsilon)}
+LC(G)+PC
\end{equation}

where $LC(G)$ is the left cut contribution and $PC$ is the pole
contribution that eventually could appear due to possible zeros of
$t_{IJ}(s)$. Now, on the right cut we have
$ImG=t_{IJ}^{(0)2}Im(1/t_{IJ})=-t_{IJ}^{(0)2}Imt_{IJ}/\mid t_{IJ}
\mid^2= -t_{IJ}^{(0)2}\sigma$. This means
that the right cut integral appearing in the dispersion relation for $G$ is the
same than that appearing in eq.10 and eq.11 for the one-loop dispersion
relation
for $t$ and hence it can be obtained from the one-loop result. The left cut
integral of the dispersion relation  in eq.12 cannot be computed exactly but we
can use the one-loop ChPT result to write  $ImG=-t_{IJ}^{(0)2}Imt_{IJ}/
\mid t_{IJ} \mid^2 \simeq
-Im t_{IJ}^{(1)}$. Besides, the substraction constants can be expanded in terms
of $
M_{\alpha}^2/F_{\beta}^2$ powers so that
$G_0=a_0-b_0+O((M_{\alpha}^2/F_{\beta}^2)^3)$, $F^2_{\alpha}G_1=a_1-b_1
+O((M_{\alpha}^2/F_{\beta}^2)^2)$, and
$F^2_{\alpha}F^2_{\beta}G_2=-b_2+O(M_{\alpha}^2/F_{\beta}^2)$. Therefore,
neglecting the the pole contribution, the dispersion relation in eq.12 can be
written approximately as:
\begin{eqnarray}
\frac{t_{IJ}^{(0)2}}{t_{IJ}}&\simeq& a_0+a_1s-b_0-b_1s-b_2s^2 \\   \nonumber
&-&\frac{s^3}{\pi}\int_{(M_{\alpha}+M_{\beta})^2}^{\infty}\frac{\sigma
t_{IJ}^{(0)2}(s')ds'}{s'^3(s'-s-i\epsilon)}+LC(G)
\end{eqnarray}
with $LC(G)$ computed with the $ImG$ approximated by $-Im t_{IJ}^{(1)}$
or in other words:
\begin{equation}
\frac{t_{IJ}^{(0)2}}{t_{IJ}}\simeq
t_{IJ}^{(0)}-t_{IJ}^{(1)}
\end{equation}
or what it is the same:
\begin{equation}
t_{IJ}\simeq
\frac{t_{IJ}^{(0)2}}{
t_{IJ}^{(0)}-t_{IJ}^{(1)} }
\end{equation}

Remarkably, to derive this result the one-loop ChPT
 approximation has been used only inside the left cut
integral but not inside the right cut integral which was computed exactly.
This is in contrast with the one-loop ChPT result in eq.10 when considered from
the point of view of the dispersion relation that $t_{IJ}$ has to fulfill. In
this case, in adddition to the above approximations used to derive eq.15, we
have done a much stronger one concerning the right cut contribution since the
$Imt_{IJ}$ was also approximated by the one-loop result. This fact is crucial
because the right cut contribution is responsible for the very strong
rescattering effects present in these reactions. As a consequence of that,
eq.15 fulfills the unitarity condition $Im t_{IJ}= \sigma \mid t_{IJ} \mid ^2$
exactly and not only perturbatively as it was the case of eq.10. Of course both
results (eq.10 and eq.15) provide the same answers at low energies but it is
expected that eq.15 will provide more realistic results at higher energies and
this is in fact what we find when comparing with the experimental data.

When zeros are present in the partial waves, they need to be included and the
final result is not so simple as in eq.15. This can be done taking into
account their contribution to eq.12 or just making a new sustraction in these
points. However, in practical cases, it can happen that the corresponding
residuous are very small and eq.15 is still a good approximation outside
the region around the position of the these zeros.  This is for instance
the case of the  $I=J=0$ and $I=2,J=0$ channels [6] which have zeros close
to $M_{\Pi}^2/2$ and  $2M_{\Pi}^2$ respectively. However, for a given
channel $IJ$ there is no way to know a priori if zeros will be present or not
since ChPT only provides a low-energy expansion of $t(s)$ and not the whole
amplitude. Nevertheless, in practice, a simple inspection of the low-energy
behaviour of the amplitudes can make sensible  the hipothesis that zeros are
not present in the low-energy region but of course this must not be necessarily
true for all the channels.

 We could also ask about the possibility of extending the above method to
higher orders of the ChPT series for instace to the two-loop computation. In
principle this can be done in a straightforward  way. We start writing a four
subtracted dispersion relation for the two-loop amplitude since it is a
third order
 polynomial modulo logarithms. As in the one-loop case, this dispersion
relation can be interpreted as an approximation to the exact amplitude
$t_{IJ}$.
Then we write a four subtracted dispersion relation for the auxiliar function $
G=t_{IJ}^{(0)2}/t_{IJ}$. The integrals of this dispersion relation can be
related modulo higher order terms to the ChPT by approaching $ImG$ by its low
energy expansion. The subtraction constants can be treated similarly
as we did in the one-loop case an finaly we arrive to:
\begin{equation}
 t_{IJ}\simeq \frac{t_{IJ}^{(0)2} } {
t_{IJ}^{(0)}-t_{IJ}^{(1)}+t_{IJ}^{(1)2}/t_{IJ}^{(0)}-t_{IJ}^{(2)}  }
\end{equation}
It is very easy to show that this amplitude, which incidentally corresponds to
the formal [1,2] Pad\'e approximant, fulfills the elastic unitarity condition
in
eq.4 but in this case, the right cut integral appearing in the dispersion
relation for $G$ cannot be computed exactly as it was in the one-loop case.
Unfortunately, as there is not any two-loop computation available, we cannot
confront the above equation with the experimental data.

 {\bf Determination of the chiral parameters and discussion:}
	First of all, we have used the set of low energy chiral parameters $L_i$
proposed in [11] for i=1,2,3 and those given in [2] for i=4..8 which will not
be
changed in this work.
	Appliying those
values to the proposed unitarized partial waves of eq.15, we have found the
appropriate qualitative behaviour for both reactions.
Moreover, we find two resonances in the partial
wave amplitudes corresponding to the $(1,1)$ and $(1/2,1)$ channels. These
resonances can be identified naturally as the $\rho$ and the $K^*$ particles.
The phase-shifts cross the $\pi/2$ value in their corresponding channels.
 In addition, in the chiral limit where an
analytical computation of the partial waves is possible, two poles are
found in he second Riemann sheet accordingly to these resonances.
 In the
other channels, where no
 physical resonances exist below $1 GeV$, they do not appear in our numerical
results. We consider this fact as a strong support for  the use of the
unitarized partial waves in eq.15.

 To fit completely the data with these formulae the next step has been to tune
the (1,1) $\pi\pi$ channel to give the correct mass for the $ \rho $ resonance.
Since this partial wave is almost only sensible to the relation $
L_3+2L^{r}_1-L^{r}_2$, then, setting the $\rho$ mass to 774 Mev means fixing a
value of this special combination of parameters.

	Nevertheless, there are still two degrees of freedom. Slightly varying the
initial values we can fit the $\pi \pi$ (2,0) and (0,0) channels, thus
obtaining $L^{r}_1$ and $L^{r}_2$. Unfortunately the experimental data
coming from these channels allow some small uncertainty in the parameters,
rending a $K^{\star}$ mass between 850 and 950 Mev. So we use the mass of this
resonance for a further parameter tuning, and finally we obtain, at the
renormalization scale $\mu=M_{\eta}=548.8 Mev$:
 \begin{equation}
L^{r}_1= 0.6 10^{-3}, L^{r}_2=1.6 10^{-3} , L_3=-3.8 10 ^{-3}
\end{equation}
 These values are well inside the errors quoted in [11]
 \begin{equation}
L^{r}_1= (0.88 \pm 0.47) 10^{-3}
, L^{r}_2=(1.61 \pm 0.38) 10^{-3}
 , L_3=(-3.62 \pm 1.31) 10^{-3}
\end{equation}

 The $K^{\star}$ mass thus
obtained is 880 Mev
	In figures 1 and 2 can be seen the results of this global fit for $\pi \pi$
scattering and those from non unitarized ChPT with the $\bar l_i$ parameters
proposed in [10].
	Figures 3 to 5 represent our global fit curves (continuous lines) for $\pi K$
scattering in three different channels.  The dashed lines are the non
unitarized ChPT predictions from the $L_i$ given in [2] and [11]. Fig.3 is
the (3/2,0) channel and figures 4 and 5 are the (1/2,1) (1/2,0)
respectively.\\

{\bf Conclusions:} The inverse amplitude method applied to the one-loop
result coming from ChPT produces a simple way to unitarize the Goldstone boson
elastic scattering amplitudes which takes into account, exactly, the
strong rescattering effects. Incidentally, this method is formally equivalent
to
the $[1,1]$ Pad\'e approximant applied to the one-loop ChPT result, provided
that the exact partial wave amplitude has no zeros in the first sheet. Note
that
 since the one-loop result is not, strictly
speaking, a polynomial,the equivalence is only formal (This is not the case if
one considers the amplitudes as polynomials in $\frac{1}{F_{\pi}^{2}}$ )

The unitarized amplitudes (with the previously fitted parameters for the
standard one-loop ChPT result) give rise to the appearence of two resonances in
the (1,1) and (1/2,1) channels that have to be understood as the $\rho$ and
$K^*$, but not any more are found in other channels where no
physical resonances exist below $1 GeV$. Therefore, the existence of these
resonances is a highly non trivial prediction of the approach followed here and
do not need to be introduced by hand in the data fit . Just tunnig slightly
only three parameters, namely $L_1$, $L_2$ and $L_3$ we are able to obtain the
right value for the masses of the $\rho$ and $K^*$  resonances. In addition we
provide a fit for six channels in a remarkable agreement with the experiment.

The results obtained in this and previous works strongly suggest that the
range of validity of the one-loop ChPT can be enlarged by a consistent
traitment
of the analiticity and unitarity constraints summarized in the dispersion
relations but a great deal of work still remains to be done in this direction.
As
a final comment, we think that the recently proposed large $N$ approximation to
ChPT (with $N$ being the number of Nambu-Godstone bosons) [25] (see also [26])
could provide a new extension of the one-loop ChPT results (see also [26] for
applications to the $\gamma\gamma \rightarrow \pi\pi$ reaction [27]).

{\bf Aknowledgements:}
This work has been partially supported by the Ministerio de Educaci\'on y
Ciencia (Spain)(CICYT AEN90-0034).

{\newpage}
{\bf Figure Captions}

	Figure 1 .- (1,1) Phase shift for $\pi \pi$ scattering. The continuous
line corresponds to our fit using eq.15. The dashed line is the result coming
from non unitarized ChPT with the $\bar l_i$ parameters proposed in [10]. The
experimental
 data comes from: $\bigcirc$ ref.[16], $\bigtriangleup$ ref.[20].

	Figure 2 .- Phase shift for $\pi \pi$ scattering. The results coming from the
fit proposed in this paper (eq.15) are the continuous line which represents the
(0,0) phase shift, and the dashed line which corresponds to that of (2,0). The
dotted  and dashed-dotted lines are the (0,0) and (2,0) phase shifts
respectively,they were obtained with non unitarized ChPT and the parameters
given in [10]. The experimental data corresponds to: $\bigtriangleup$ ref.[12],
$\bigcirc$ ref.[13],$\Box$ ref.[14],$\diamondsuit$ ref.[15],
$\bigtriangledown$ ref.[16],$\star$ ref.[17],$\times$ ref.[18],$\bullet$
ref.[19].

	Figure 3.- Phase shift of the (3/2,0) channel for $\pi K$ scattering. The
continuous line is the result of the inverse amplitud method (eq.15) with the
parameters proposed in this paper, whereas the dashed line is non unitarized
ChPT with the parameters proposed in [2] and [11]. Data corresponds to ref.[21]

	Figure 4.- The same as figure 3 but for the (1/2,1) partial wave. Data comes
from: $\bigtriangleup$ ref.[21] and $\star$ ref.[24].

	Figure 5.- As figure 3 but for the (1/2,0) channel. The experimental data
corresponds to: $\bigtriangleup$ ref.[21],$\diamondsuit$ ref.[22], and
$\star$ to ref.[23].

 {\bf D41} (1990) 3324
\bibitem{27} J. Bijnens and F. Cornet, {\em Nuc. Phys.} {\bf
B296}(1988)557 \\ J.F. Donoghue, B.R. Holstein and Y.C.
Lin, {\em Phys. Rev.} {\bf D37}(1988)2423\\ A. Dobado and J.R. Pel\'aez,
preprint FT/UCM/9/92. To appear in {\em Z. Phys.}
\newpage
\end{document}